\documentclass[useAMS,usenatbib]{mn2e}

\usepackage[dvips]{graphicx}
\usepackage[english]{babel}

\title[Models of cuspy triaxial stellar systems. II. Regular orbits]
{Models of cuspy triaxial stellar systems. II. Regular orbits}
\author[J. C. Muzzio, H.D. Navone and A. F. Zorzi]{J. C. Muzzio$^{1}$\thanks{E-mail: 
jcmuzzio@fcaglp.unlp.edu.ar (JCM); navone@ifir-conicet.gov.ar; 
azorzi@fceia.unr.edu.ar (AFZ)}, H. D. Navone $^{2}$and A. F. 
Zorzi$^{2}$\\
$^{1}$Instituto de Astrof\'{i}sica de La Plata (CONICET La Plata--UNLP) 
and Facultad de Ciencias Astron\'omicas y Geof\'{i}sicas, \\
Universidad Nacional de La Plata, La Plata, Argentina\\
$^{2}$Instituto de F\'{i}sica de Rosario (CONICET--UNR) and Facultad de 
Ciencias Exactas, Ingenier\'{i}a y Agrimensura,\\
Universidad Nacional de Rosario, Rosario, Argentina\\}
\begin{document}

\date{XXX}

\pagerange{\pageref{firstpage}--\pageref{lastpage}} \pubyear{XXX}

\maketitle

\label{firstpage}

\begin{abstract}
In the first paper of this series we used the N--body method to build a dozen
cuspy  ($\gamma \simeq 1$) triaxial models of stellar systems, and we showed that
they were highly stable over time intervals of the order of a Hubble time,
even though they had very large fractions of chaotic orbits (more than 85 per cent in
some cases). The models were  grouped in four sets, each one comprising models
morphologically resembling E2, E3, E4 and E5 galaxies, respectively. The three models
within each set, although different, had the same global properties and were
statistically equivalent. In the present paper we use frequency analysis to classify
the regular orbits of those models. The bulk of those orbits are short axis tubes (SATs),
with a significant fraction of long axis tubes (LATs) in the E2 models that decreases in
the E3 and E4 models to become negligibly small in the E5 models. Most of the LATs in
the E2 and E3 models are outer LATs, but the situation reverses in the E4 and E5 models
where the few LATs are mainly inner LATs. As could be expected for cuspy models, most of
the boxes are resonant orbits, i.e., boxlets. Nevertheless, only the (x, y) fishes of
models E3 and E4 amount to about 10 per cent of the regular orbits, with most of the
fractions of the other boxlets being of the order of 1 per cent or less.

\end{abstract}


\section{Introduction}

The observational evidence, both statistical \citep{R1996} and on individual galaxies
\citep{SEPB2004}, indicates that at least some elliptical galaxies are triaxial and not merely
rotationally symmetric. Besides, their surface brightness increases towards the center, forming
a cusp \citep{Cea1993,MSZ1995} that reveals a mass concentration, and perhaps the presence of
a black hole, there. Thus, the need of triaxial and cuspy models to represent elliptical
galaxies seems to be warranted.

Triaxial models of stellar systems with smooth cores harbour four major families of regular
orbits: boxes, short axis tubes (SATs hereafter) and inner and outer long axis tubes (ILATs and
OLATs, respectively, hereafter); see, e.g., \citet{dZ1985} or \citet{S1987}. Significant resonant
orbit families, called boxlets, were found in the singular logarithmic potential \citep{B1982,MES1989}
and, in general, they tend to replace the box orbits in models with central cusps. Although chaotic
orbits were originally thought to make only a minor contribution to the orbital content of triaxial
models, they were later recognized to arise naturally in those models, specially in cuspy ones
\citep{KS2003}.

In a recent paper \citep{ZM2012}, herein Paper I, we have presented self--consistent models of
cuspy triaxial stellar systems obtained using the N--body method. The models are morphologically
similar to elliptical galaxies of Hubble types E2 through E5, with de Vaucouleurs density profiles.
We showed that they were very stable over time intervals of the order of one Hubble time, even
though they contained extremely high fractions of chaotic orbits (higher than 85 per cent in half
of the models). Thus, the usual idea that the regular orbits provide the backbone of stellar
systems is in doubt for these models. In the present paper we use frequency analysis techniques
to classify the regular orbits found in our previous investigation, in order to determine which
kinds of regular orbits are present in the models of Paper I and in which proportions they appear.

In the next section we describe the models and the classification technique.
Section 3 presents our results and Section 4 summarizes our conclusions.

\section{Models and techniques}

\subsection{The models}

A detailed explanation of how the models were built is given in Paper I. Briefly, the recipe
of \citet{AM1990} was used, randomly creating a spherical distribution of $10^6$ particles with a
density distribution inversely proportional to the distance to the center and a Gaussian velocity
distribution, and following its collapse with the code of \citet{HO1992};
the resulting triaxial system is a consequence of the radial orbit instability. The gravitational
constant, $G$, the radius of the sphere and the total mass are all set equal to $1$. The models
were rotated to have the major, intermediate and minor axes of their moment of inertia
tensor aligned with the x, y and z coordinate axes, respectively (the corresponding velocity
components are dubbed u, v and w, respectively, hereafter). Particles with energies close
to, or larger than, zero were eliminated and the models were allowed to relax
to make sure they had reached equilibrium. Tables 1 and 2 of Paper I give the main properties of the
models and, in particular, crossing times ($T_{cr}$ hereafter) are of the order of 0.5 time units
(t.u. hereafter); the Hubble time was found to be of the order of 100 t.u., or about $200 T_{cr}$.
The major, intermediate and minor axes of the models ($a, b$ and $c$, respectively)
were obtained from the mean square values of the $x, y$ and $z$ coordinates, respectively, taking the
20, 40, ..., 100 per cent most tightly bound particles. The major semiaxes and the semiaxes ratios
change with the orbital energy limit and are given in Table 2 of Paper I. The Hubble type was estimated
from the $c/a$ ratio of the 80 most tightly bound particles, but it should be noticed that triaxiality
($T=(a^2-b^2)/(a^2-c^2)$), given in Table 1 of Paper I, goes from about 0.73 (i.e., not too far from
being prolate) for the E2 models to about 0.47 (i.e., close to maximum triaxiality) for the E5 ones. To
aid the reader, we give the $a$, $b/a$ $c/a$ values also in Table \ref{tab2} of the present paper but,
in this case, the results for the three models of each group were bunched together and they were computed
in energy bins, instead of groups of the 20, 40, ..., 100 per cent most bounded particles as in Paper I.
The density distributions follow the de Vaucouleurs law and all the models are cuspy, with central
densities proportional to $r^{-\gamma}$ and $\gamma \simeq 1$. The twelve models are divided in four
groups, E2, E3, E4 and E5, named after the elliptical galaxy classes that correspond to their axial
ratios. Three models ($a, b$ and $c$) were created for each group using different seed numbers for
the random number generator, so that they are statistically equivalent; in fact, as shown in Paper I,
the macroscopic properties of the three models in each group turned out to be essentially the same.

We showed in Paper I that all the models are highly stable over time intervals of the order of a
Hubble time. Their orbital structure is dominated by chaotic orbits, with regular orbits amounting
to little more than 20 per cent in models E2 and E5 and to less than 15 per cent in models E3 and E4.

\subsection{Frequency analysis and regular orbit classification}

As in our previous papers \citep{Muz2006, AMNZ2007, MNZ2009}, the modified Fourier transform code of
\citet{SN1997} (a copy can be obtained at $www.boulder.swri.edu/{\sim}davidn$) was used for the frequency
analysis. In Paper I the positions and velocities of about 5,000 bodies were ramdomly selected from each
model and adopted as initial conditions for the computation of the Lyapunov exponents, which
allowed us to separate the regular from the chaotic orbits. For the 9,899 orbits deemed there as regular,
we adopted here those same initial positions and velocities to compute the corresponding orbits and we
obtained the fundamental frequencies for each coordinate, $F_{x}$, $F_{y}$ and $F_{z}$, performing
the frequency analysis on the complex variables $x+iu, y+iv$ and $z+iw$, respectively; these were
derived from 8,192 points equally spaced in time obtained integrating the regular orbits over 300
radial periods. In this way, as indicated by \citet{Muz2006}, frequencies can be obtained with a
precision better than $10^{-9}$ for isolated lines; nevertheless,  the precision is much lower
when there are nearby lines and the practical limit of $2 \times 10^{-4}$ for the precision,
adopted in our previous works, is also used here.

The orbits were then classified as boxes and boxlets (BBLs hereafter), SATs, ILATs and OLATs
using the method of \citet{KV2005}, with the
improvements introduced by \citet{Muz2006}, \citet{AMNZ2007} and \citet{MNZ2009}. The original
method took the frequency of the largest amplitude component in each coordinate as the fundamental
frequency for that coordinate but, as shown by \citet{BS1982} and \citet{Muz2006}, respectively, the
libration of some orbits and the extreme elongation of others makes necessary to adopt other
frequencies as the fundamental ones, so that some of the improvements deal with those cases.
Besides, \citet{AMNZ2007} showed that one has to take into account the energy of the orbit, in
addition to the $F_{x}/F_{z}$ ratio used by \citet{KV2005}, to separate ILATs from OLATs and that
is another improvement of the original method. Finally, we searched for resonances among the
fundamental frequencies of the BBLs in order to separate the boxes from the boxlets.

\section{Results and analysis}

\begin{figure*}
\vspace{20pt}
\centering
\begin{minipage}{160mm}
\includegraphics[width=7.5truecm]{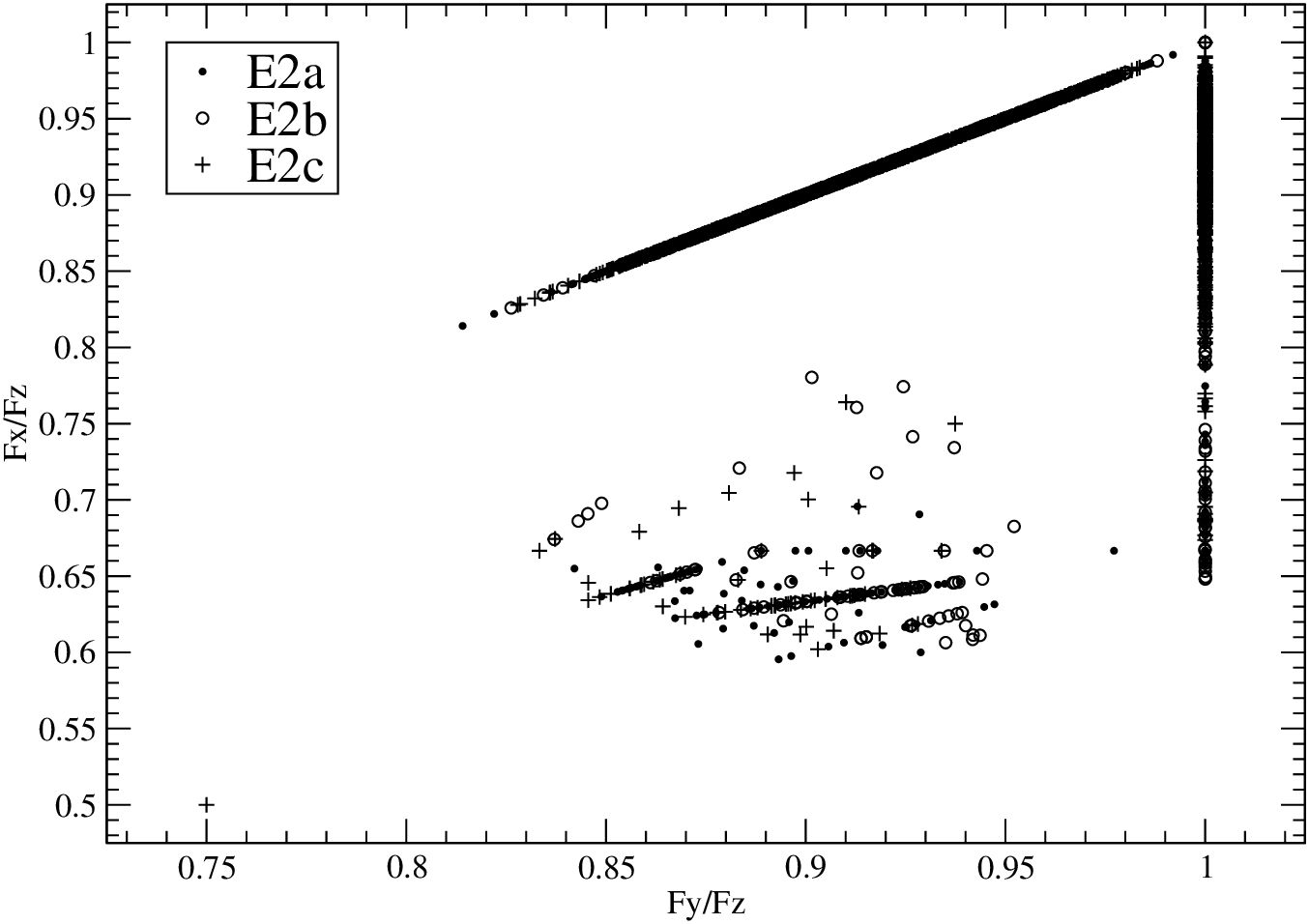}~\hfill\includegraphics[width=7.5truecm]{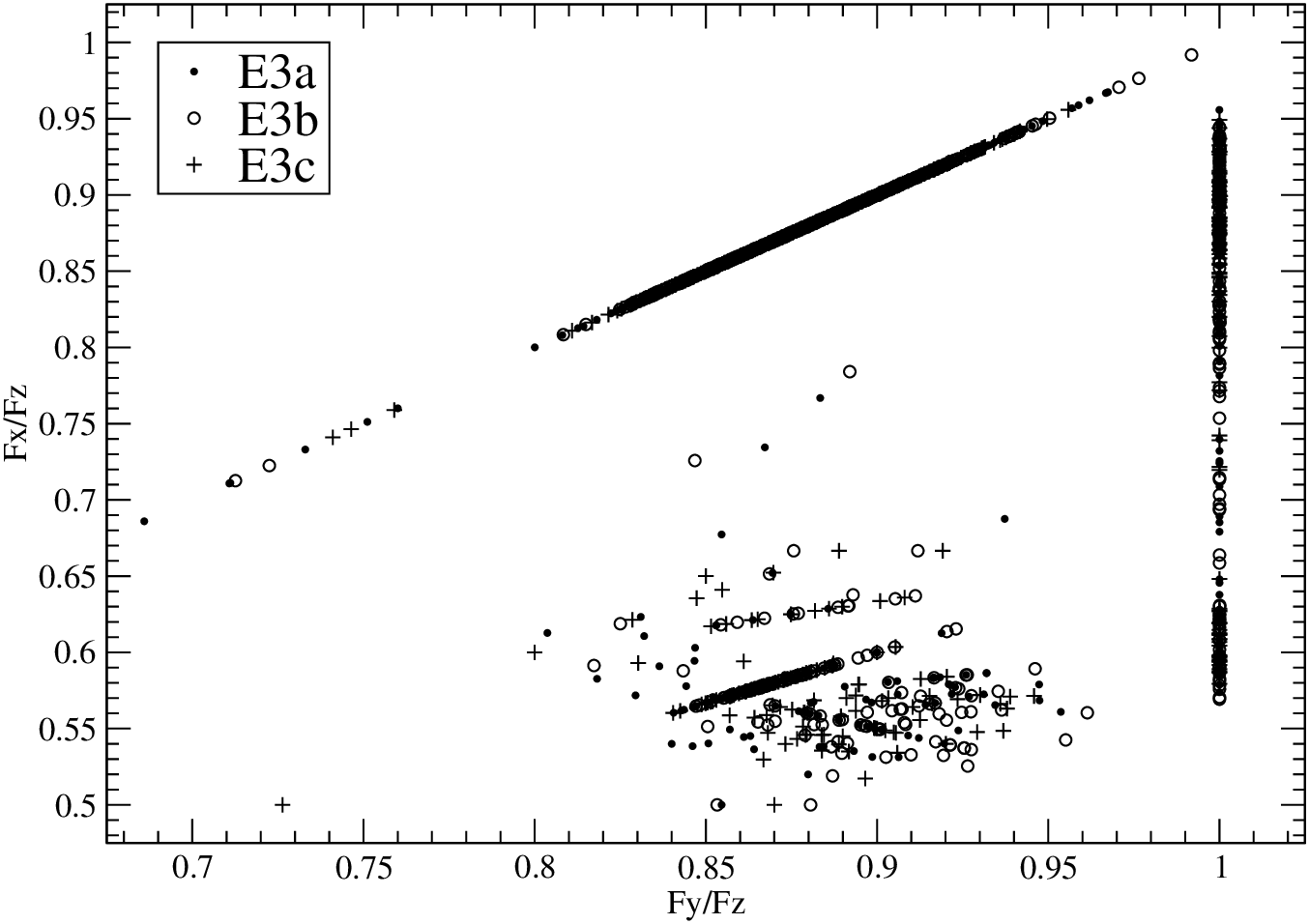}
\vskip 10.5mm
\includegraphics[width=7.5truecm]{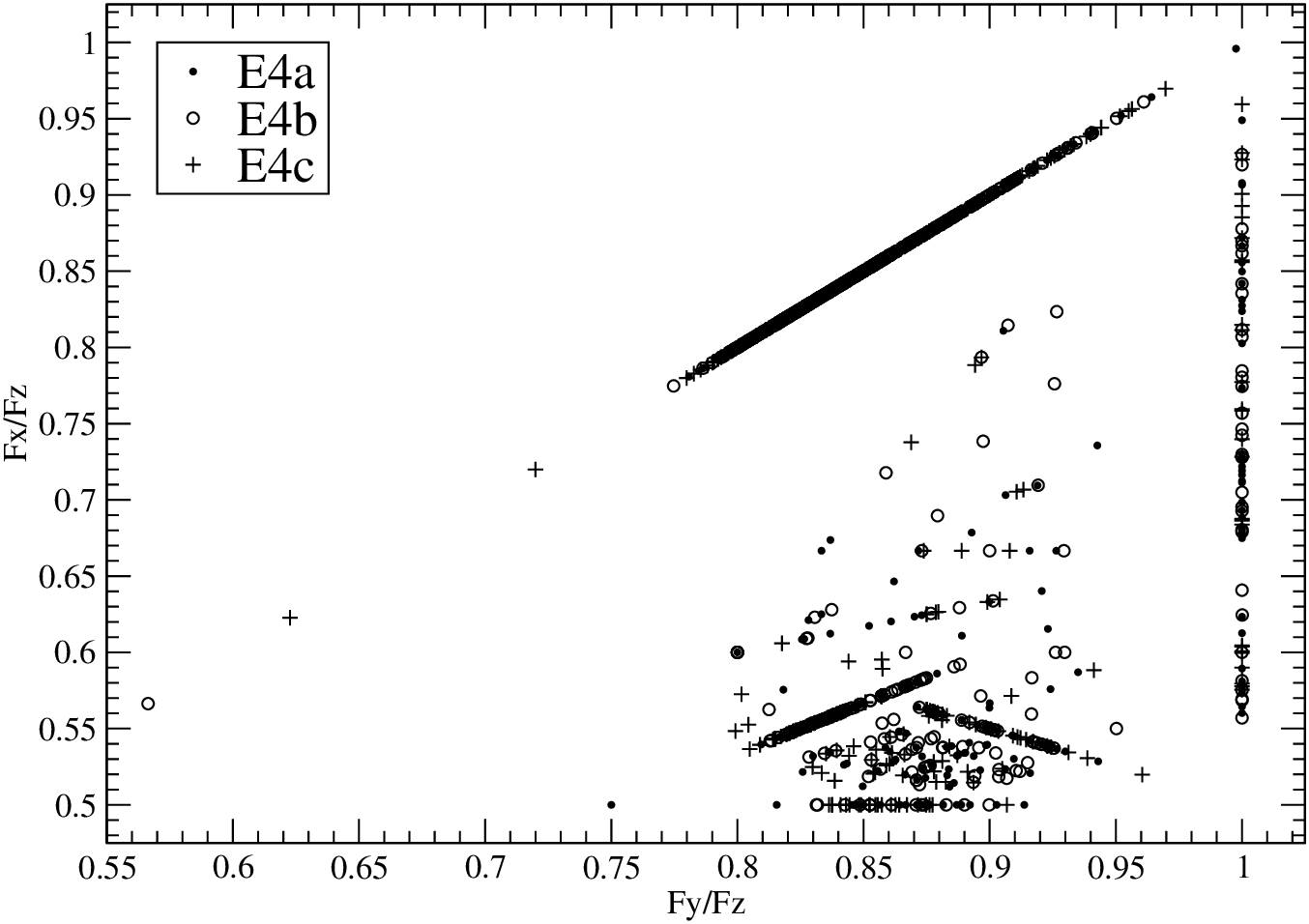}~\hfill\includegraphics[width=7.5truecm]{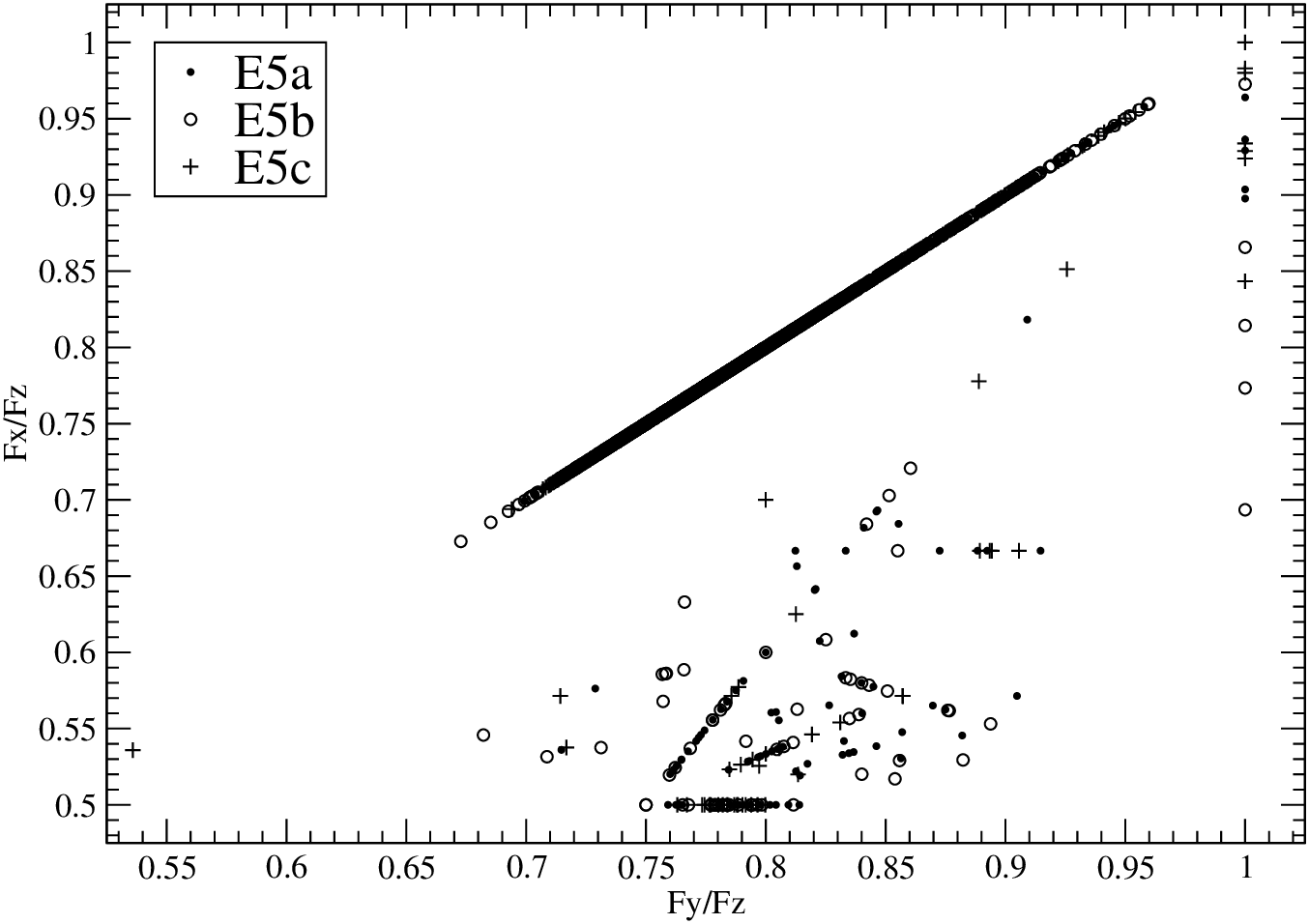}
\caption{Frequency maps for models E2 (top left), E3 (top right), E4 (bottom left) and E5 (bottom
right). Please note that the horizontal scale is different for each model.}
\label{fig01}
\end{minipage}
\end{figure*}

Of the 9,899 orbits regarded as regular in Paper I, 180 yielded anomalous values of their fundamental
frequencies, i.e., frequencies that did not obey that $F_{x} \le F_{y} \le F_{z}$ or whose
$F_{y}/F_{z}$ or $F_{x}/F_{z}$ ratios placed them at odd locations on the frequency map. Visual
inspection of their spectra showed that most of them were typical of chaotic orbits, with
lines of similar frequencies and amplitudes. We checked that possibility obtaining the finite time
Lyapunov characteristic numbers (FT-LCNs hereafter, see Paper I for details) of those orbits
using an integration time of 100,000 t.u., i.e. ten times longer than that used in
Paper I. The limiting value to separate regular from chaotic orbits with the longer integration
time was found to be $0.00020 (t.u.)^{-1}$, while in Paper I it was $0.00180 (t.u.)^{-1}$.
Of the 180 suspicious orbits, 164 turned out to be actually chaotic and were
eliminated from the subsequent analysis.  Plots of the remaining 16 orbits, showed them to be normal
tubes, but most had highly elongated orbits so that, as indicated by \citet{Muz2006}, their fundamental
frequencies should not be taken as those corresponding to the largest amplitude. In fact, relaxing
the condition that allows to adopt a frequency different from that of largest amplitude as the
fundamental frequency, let our classification code to automatically select the right frequencies,
but this procedure is risky and may spoil the classification of other orbits, so that we preferred to
select the fundamental frequencies for those few orbits from visual inspection of their frequency spectra.

\begin{figure}
\vspace{20pt}
\includegraphics[width=84mm]{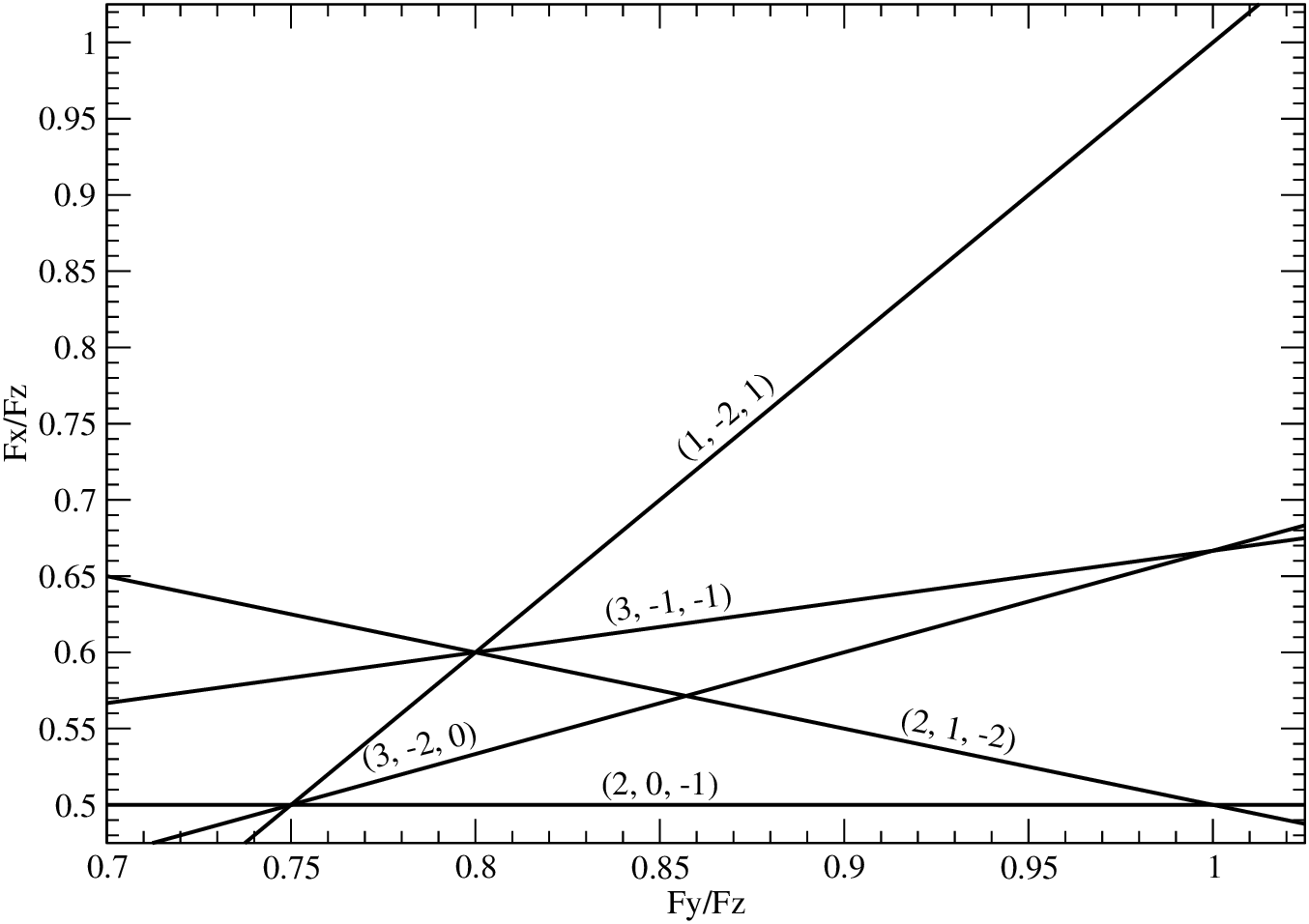}
\caption{Main resonances in the frequency maps.}
\label{fig02}
\end{figure}

Figure \ref{fig01} presents the frequency maps for each group and, within each panel,
different symbols are used for each one of the three statistically equivalent
models. The plots are bound by the $F_{x}/F_{z} = F_{y}/F_{z}$ correlation, corresponding to
the SATs, and the $ F_{y}/F_{z} = 1$ correlation, corresponding to the LATs. Besides, several
other correlations, corresponding to different boxlets, are evident on the plots. Within each
panel there is generally good agreement among the results from the different models within each group.
To aid the reader, the main resonances that stand out in Figure \ref{fig01} are
identified in Figure 2.

\begin{table*}
\centering
\begin{minipage}{140mm}
\caption{Orbital classification results}
\label{tab1}
\begin{tabular}{@{}lcccccc@{}}
\hline
Model & Total & Chaotic & BBL & SAT & ILAT & OLAT \\
$ $ & $ $ &  (per cent) & (per cent) & (per cent) & (per cent) & (per cent) \\ 
\hline 
E2a & 1113 & 0.54 $\pm$ 0.22 & 9.88 $\pm$ 0.89 & 54.72 $\pm$ 1.49 & 3.14 $\pm$ 0.52 & 31.72 $\pm$ 1.39 \\ 
E2b & 1057 & 0.28 $\pm$ 0.16 & 7.66 $\pm$ 0.82 & 51.66 $\pm$ 1.54 & 3.69 $\pm$ 0.58 & 36.71 $\pm$ 1.48 \\ 
E2c & 1101 & 0.54 $\pm$ 0.22 & 6.63 $\pm$ 0.75 & 59.58 $\pm$ 1.48 & 2.27 $\pm$ 0.45 & 30.97 $\pm$ 1.39 \\ 
\hline
E3a & 696 & 6.03 $\pm$ 0.90 & 20.26 $\pm$ 1.52 & 57.76 $\pm$ 1.87 & 7.04 $\pm$ 0.97 & 8.91 $\pm$ 1.08 \\ 
E3b & 687 & 3.20 $\pm$ 0.67 & 23.00 $\pm$ 1.61 & 53.86 $\pm$ 1.90 & 6.84 $\pm$ 0.96 & 13.10 $\pm$ 1.29 \\ 
E3c & 520 & 4.04 $\pm$ 0.86 & 25.58 $\pm$ 1.91 & 53.85 $\pm$ 2.18 & 6.54 $\pm$ 1.08 & 10.00 $\pm$ 1.32 \\ 
\hline
E4a & 594 & 2.02 $\pm$ 0.58 & 32.15 $\pm$ 1.92 & 59.76 $\pm$ 2.01 & 5.05 $\pm$ 0.90 & 1.01 $\pm$ 0.41 \\ 
E4b & 576 & 2.78 $\pm$ 0.68 & 31.42 $\pm$ 1.93 & 60.24 $\pm$ 2.04 & 4.69 $\pm$ 0.88 & 0.87 $\pm$ 0.39 \\ 
E4c & 575 & 2.09 $\pm$ 0.60 & 26.78 $\pm$ 1.85 & 66.43 $\pm$ 1.97 & 4.00 $\pm$ 0.82 & 0.70 $\pm$ 0.35 \\ 
\hline
E5a & 984 & 0.41 $\pm$ 0.20 & 8.03 $\pm$ 0.87 & 91.06 $\pm$ 0.91 & 0.51 $\pm$ 0.23 & 0.00 $\pm$ 0.00 \\ 
E5b & 1047 & 1.15 $\pm$ 0.33 & 6.21 $\pm$ 0.75 & 92.17 $\pm$ 0.83 & 0.48 $\pm$ 0.21 & 0.00 $\pm$ 0.00 \\
E5c & 949 & 0.84 $\pm$ 0.30 & 4.32 $\pm$ 0.66 & 94.10 $\pm$ 0.76 & 0.63 $\pm$ 0.26 & 0.11 $\pm$ 0.11 \\ 
\hline
\end{tabular}
\end{minipage}
\end{table*}

Except for the 16 orbits mentioned before, whose fundamental frequencies were obtained from
visual inspection of their spectra, all the others were automatically classified by our code.
LATs were segregated into ILATs and OLATs using $F_{x}/F_{z}$ versus orbital energy plots, as
explained by \citet{AMNZ2007}. Table \ref{tab1} summarizes the classification results. It gives,
for each model, the total number of regular orbits found in Paper I, and the percentages of them that
turned out to be chaotic with the longer integration interval, and that were classified as BBLs, SATs,
ILATs and OLATs; the statistical errors have been estimated from the binomial distribution, as
in our previous papers.

\begin{table*}
\centering
\begin{minipage}{175mm}
\caption{Orbital classification results per energy bin}
\label{tab2}
\begin{tabular}{@{}lcccccccccc@{}}
\hline
Model & Energy Bin & a & b/a & c/a & Total & Chaotic & BBL & SAT & ILAT & OLAT \\
$ $ & (per cent) & $ $ & $ $ & $ $ & $ $ & (per cent) & (per cent) & (per cent) & (per cent) & (per cent) \\ 
\hline 
E2 & 0-20 & 0.068 & 0.754 & 0.601 & 578 & 0.00 $\pm$ 0.00 & 14.53 $\pm$ 1.47 & 74.39 $\pm$ 1.82 & 10.90 $\pm$ 1.30 & 0.17 $\pm$ 0.17 \\ 
$ $ & 20-40 & 0.151 & 0.795 & 0.704 & 325 & 0.00 $\pm$ 0.00 & 15.38 $\pm$ 2.00 & 78.15 $\pm$ 2.29 & 3.38 $\pm$ 1.00 & 3.08 $\pm$ 0.96 \\ 
$ $ & 40-60 & 0.251 & 0.853 & 0.799 & 455 & 0.00 $\pm$ 0.00 & 7.91 $\pm$ 1.27 & 55.16 $\pm$ 2.33 & 0.88 $\pm$ 0.44 & 36.04 $\pm$ 2.25 \\ 
$ $ & 60-80 & 0.512 & 0.891 & 0.847 & 704 & 0.00 $\pm$ 0.00 & 4.69 $\pm$ 0.80 & 51.56 $\pm$ 1.88 & 0.57 $\pm$ 0.28 & 43.18 $\pm$ 1.87 \\ 
$ $ & 80-100 & 1.145 & 0.945 & 0.930 & 1209 & 1.24 $\pm$ 0.32 & 5.05 $\pm$ 0.63 & 42.43 $\pm$ 1.42 & 1.41 $\pm$ 0.34 & 49.88 $\pm$ 1.44 \\ 
\hline
E3 & 0-20 & 0.064 & 0.725 & 0.570 & 454 & 0.00 $\pm$ 0.00 & 11.89 $\pm$ 1.52 & 78.19 $\pm$ 1.94 & 9.91 $\pm$ 1.40 & 0.00 $\pm$ 0.00 \\ 
$ $ & 20-40 & 0.151 & 0.711 & 0.557 & 223 & 0.00 $\pm$ 0.00 & 27.80 $\pm$ 3.00 & 69.51 $\pm$ 3.08 & 1.79 $\pm$ 0.89 &0.90 $\pm$ 0.63 \\ 
$ $ & 40-60 & 0.234 & 0.800 & 0.671 & 191 & 0.00 $\pm$ 0.00 & 22.51 $\pm$ 3.02 & 69.11 $\pm$ 3.34 & 2-09 $\pm$ 1.04 & 6.28 $\pm$ 1.76 \\ 
$ $ & 60-80 & 0.515 & 0.819 & 0.707 & 301 & 0.00 $\pm$ 0.00 & 19.60 $\pm$ 2.29 & 61.46 $\pm$ 2.81 & 2.33 $\pm$ 0.87 & 16.61 $\pm$ 2.15 \\ 
$ $ & 80-100 & 2.449 & 0.850 & 0.813 & 734 & 11.58 $\pm$ 1.18 & 29.16 $\pm$ 1.68 & 30.65 $\pm$ 1.70 & 9.54 $\pm$ 1.08 & 19.07 $\pm$ 1.45 \\ 
\hline
E4 & 0-20 & 0.056 & 0.736 & 0.582 & 421 & 0.00 $\pm$ 0.00 & 18.53 $\pm$ 1.89 & 69.36 $\pm$ 2.25 & 12.11 $\pm$ 1.59 & 0.00 $\pm$ 0.00 \\ 
$ $ & 20-40 & 0.135 & 0.694 & 0.504 & 256 & 0.00 $\pm$ 0.00 & 28.12 $\pm$ 2.81 & 69.92 $\pm$ 2.87 & 1.56 $\pm$ 0.78 & 0.39 $\pm$ 0.39 \\ 
$ $ & 40-60 & 0.218 & 0.750 & 0.557 & 297 & 0.00 $\pm$ 0.00 & 27.95 $\pm$ 2.60 & 69.70 $\pm$ 2.67 & 2.36 $\pm$ 0.88 & 0.00 $\pm$ 0.00 \\ 
$ $ & 60-80 & 0.383 & 0.780 & 0.595 & 409 & 0.00 $\pm$ 0.00 & 21.27 $\pm$ 2.02 & 76.53 $\pm$ 2.10 & 0.98 $\pm$ 0.49 & 1.22 $\pm$ 0.54 \\ 
$ $ & 80-100 & 2.557 & 0.789 & 0.701 & 362 & 11.05 $\pm$ 1.65 & 56.91 $\pm$ 2.60 & 25.69 $\pm$ 2.30 & 3.87 $\pm$ 1.01 & 2.49 $\pm$ 0.82 \\ 
\hline
E5 & 0-20 & 0.051 & 0.839 & 0.579 & 676 & 0.00 $\pm$ 0.00 & 5.03 $\pm$ 0.84 & 92.75 $\pm$ 1.00 & 2.07 $\pm$ 0.55 & 0.15 $\pm$ 0.15 \\ 
$ $ & 20-40 & 0.124 & 0.811 & 0.497 & 602 & 0.00 $\pm$ 0.00 & 2.49 $\pm$ 0.64 & 97.18 $\pm$ 0.68 & 0.33 $\pm$ 0.23 & 0.00 $\pm$ 0.00 \\
$ $ & 40-60 & 0.216 & 0.807 & 0.501 & 483 & 0.00 $\pm$ 0.00 & 5.59 $\pm$ 1.05 & 94.41 $\pm$ 1.05 & 0.00 $\pm$ 0.00 & 0.00 $\pm$ 0.00 \\ 
$ $ & 60-80 & 0.358 & 0.808 & 0.512 & 537 & 0.00 $\pm$ 0.00 & 6.52 $\pm$ 1.07 & 93.48 $\pm$ 1.07 & 0.00 $\pm$ 0.00 & 0.00 $\pm$ 0.00 \\ 
$ $ & 80-100 & 0.993 & 0.899 & 0.573 & 682 & 3.52 $\pm$ 0.71 & 10.85 $\pm$ 1.19 & 85.63 $\pm$ 1.34 & 0.00 $\pm$ 0.00 & 0.00 $\pm$ 0.00 \\ 
\hline
\end{tabular}
\end{minipage}
\end{table*}

Table \ref{tab2} presents the results of the classification for the orbits grouped in
orbital energy bins and, since the numbers are small, the results of the three models of each group have been
bunched together. The first column gives the group and the second one gives the energy range of the bin; columns
three through five give the major semiaxis and the semiaxes ratios corresponding to the bin; the other columns
are as in Table \ref{tab1} except that the rows correspond to the energy bins rather that to the different
models.

\begin{table*}
\centering
\begin{minipage}{140mm}
\caption{Percentages of boxes and boxlets.} 
\label{tab3}
\begin{tabular}{@{}lcccc@{}}
\hline
Type & E2 & E3 & E4 & E5 \\ 
 & (per cent) & (per cent) & (per cent) & (per cent) \\ 
\hline 
 BBL & 8.07 $\pm$ 0.48 & 22.70 $\pm$ 0.96 & 30.14 $\pm$ 1.10 & 6.21 $\pm$ 0.44 \\ 
 Boxes($\leq 5$) & 0.79 $\pm$ 0.16 & 3.89 $\pm$ 0.44 & 2.98 $\pm$ 0.41 & 0.23 $\pm$ 0.09 \\ 
 (1,-2,1) & 0.15 $\pm$ 0.07 & 0.16 $\pm$ 0.09 & 0.46 $\pm$ 0.16 & 1.07 $\pm$ 0.19 \\ 
 (2,0,-1) & 0.00 $\pm$ 0.00 & 0.26 $\pm$ 0.12 & 3.09 $\pm$ 0.41 & 1.71 $\pm$ 0.24 \\
 (2,1,-2) & 0.00 $\pm$ 0.00 & 2.31 $\pm$ 0.34 & 3.78 $\pm$ 0.46 & 0.40 $\pm$ 0.12 \\
 (3,-2,0) & 0.76 $\pm$ 0.15 & 8.57 $\pm$ 0.64 & 10.43 $\pm$ 0.73 & 0.64 $\pm$ 0.15 \\
 (3,-1,-1) & 3.15 $\pm$ 0.31 & 1.26 $\pm$ 0.26 & 1.09 $\pm$ 0.25 & 0.27 $\pm$ 0.09 \\
 (3,-3,1) & 0.03 $\pm$ 0.03 & 0.47 $\pm$ 0.16 & 0.86 $\pm$ 0.22 & 0.00 $\pm$ 0.00 \\
 (4,-3,0) & 1.47 $\pm$ 0.21 & 0.32 $\pm$ 0.13 & 0.23 $\pm$ 0.11 & 0.13 $\pm$ 0.07 \\
 (5,-3,0) & 0.00 $\pm$ 0.00 & 0.16 $\pm$ 0.09 & 0.97 $\pm$ 0.24 & 0.00 $\pm$ 0.00 \\ 
 (5,-2,-1) & 0.00 $\pm$ 0.00 & 1.05 $\pm$ 0.23 & 0.69 $\pm$ 0.20 & 0.13 $\pm$ 0.07 \\ 
 Other ($\leq 5$) & 1.22 $\pm$ 0.19 & 1.84 $\pm$ 0.31 & 2.69 $\pm$ 0.39 & 0.81 $\pm$ 0.16 \\ 
 2 resonances ($\leq 5$) & 0.49 $\pm$ 0.12 & 2.42 $\pm$ 0.35 & 2.87 $\pm$ 0.40 & 0.81 $\pm$ 0.16 \\ 
\hline
 Boxes($\leq 10$) & 0.21 $\pm$ 0.08 & 1.58 $\pm$ 0.29 & 1.03 $\pm$ 0.24 & 0.07 $\pm$ 0.05 \\ 
\hline
\end{tabular} 
\end{minipage}
\end{table*}

Although we had found before \citep{MNZ2009} that some orbits deemed to be regular from their FT-LCNs
turned out to be chaotic during the orbital classification, the percentages shown in Tables
\ref{tab1} and \ref{tab2} are worrisome. As a check, for all the 576 orbits of model E4b classified as regular
in Paper I, we obtained the FT-LCNs using the ten times longer interval. For 172 orbits the new FT-LCNs turned
out to be larger than the new limiting value of $0.00020 (t.u.)^{-1}$, i.e., they should be regarded as
chaotic. Nevertheless, only 33 of those had FT-LCNs larger than the limiting value of $0.00180 (t.u.)^{-1}$
of Paper I. In other words, only $5.73 \pm 0.31$ per cent of the "regular" orbits of Paper I turned
out to be sticky orbits that actually had FT-LCNs larger than the limit adopted there. In addition, another
$24.13 \pm 0.18$ per cent were weakly chaotic orbits whose FT-LCNs were simply below the detection limit used
in Paper I. Considering this result, the percentages of chaotic orbits of Table
\ref{tab1} are not very high, in all likelihood because they were found rather accidentally from oddities of
the frequency analysis, but one should remember that they are just the tip of the iceberg and that any
sample of regular obits obtained using chaos indicators, such as the FT-LCNs, is bound to include
a substantial amount of chaotic orbits as well. Fortunately, at least in our case, most of these are
weakly chaotic orbits whose behaviour is not too different from that of regular orbits \citep{KV2005},
with sticky orbits that might exhibit a wilder behaviour being a minority only.
The bulk of the sticky orbits (26 of them) were concentrated in the highest 20 per
cent energy bin, while the numbers of weakly chaotic orbits raised steadily from the lowest (7 orbits) through
the highest (56 orbits) energy bins.

\begin{figure}
\vspace{20pt}
\centering
\includegraphics[width=8.4truecm]{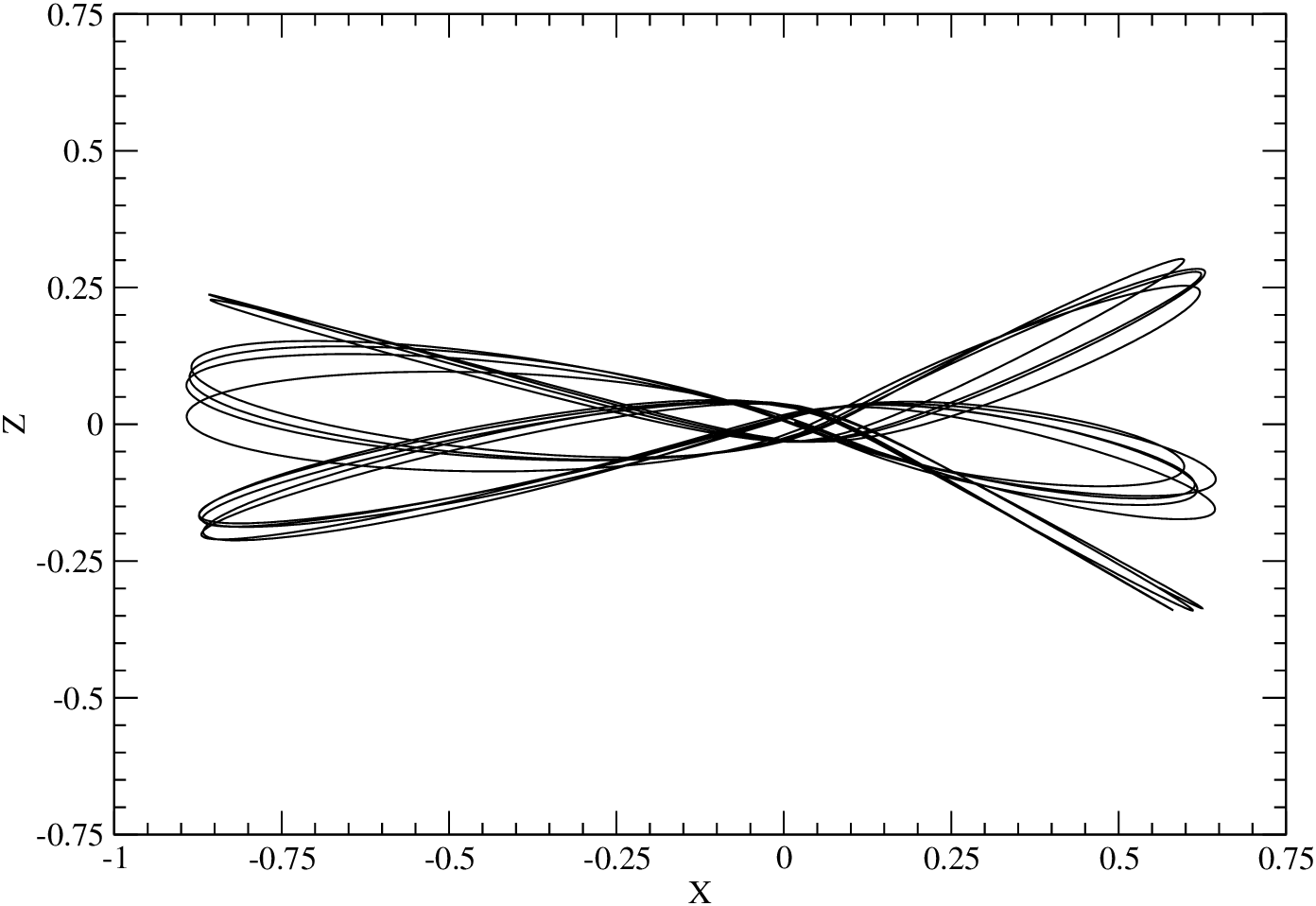}
\vskip 10.5mm
\includegraphics[width=8.4truecm]{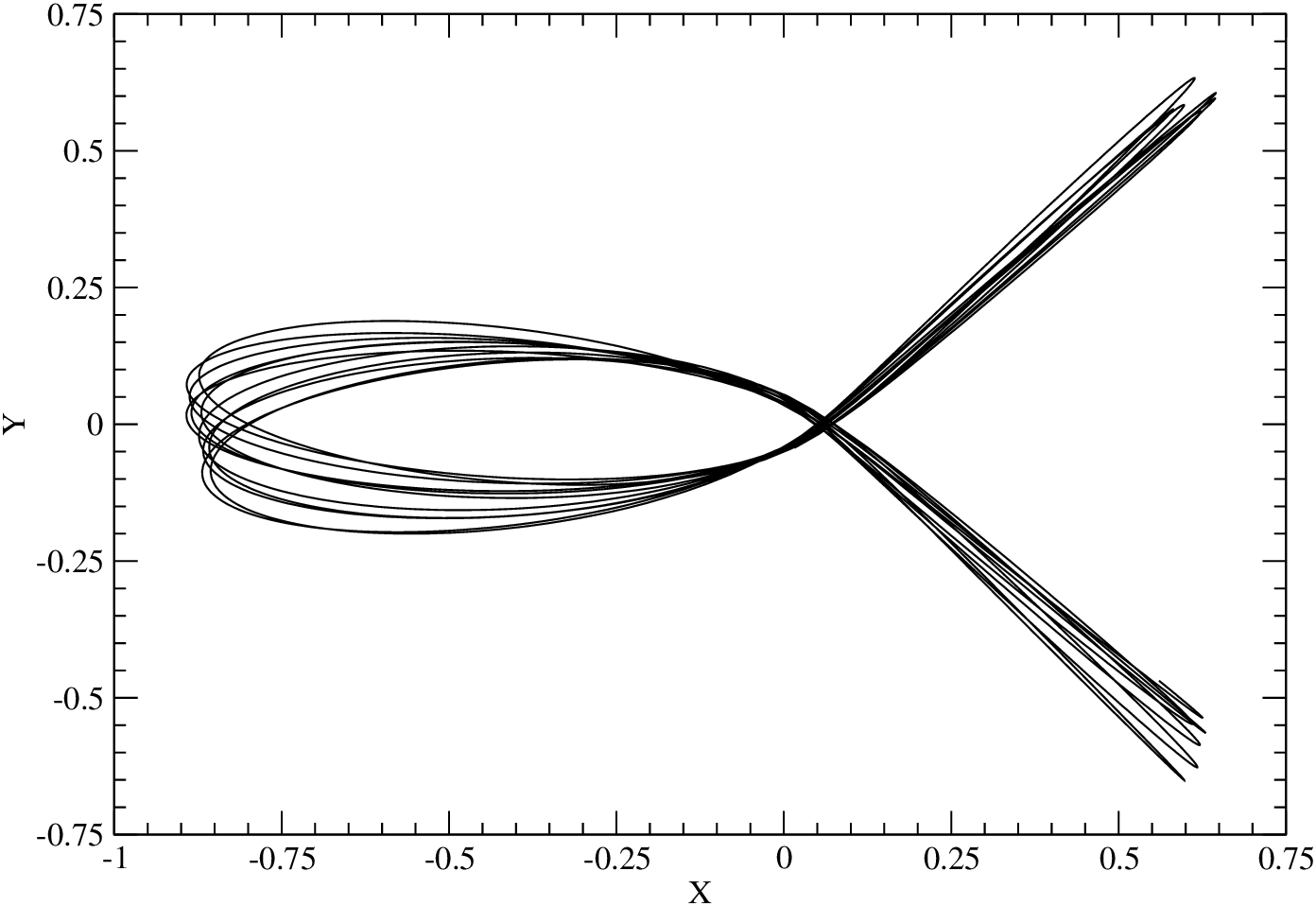}
\caption{Projections on the x-z (top) and x-y (bottom) planes of resonant orbit 1848 from
model E4a. The orbit obeys a (3,-2,0) resonance, i.e., it is a fish.}
\label{fig03}
\end{figure}

Since the BBLs include both true boxes and resonant boxes (boxlets), it is important to segregate ones
from the others. Thus, we searched for resonances obeying the relationship:

\begin{equation}
l F_{x} + m F_{y} + n F_{z} = 0,
\end{equation}

with $l, m$ and $n$ integers not all equal to zero. Since our computed frequencies are not exact,
the above relationship can be fullfilled only approximately, and it is rather risky
to search for resonances that involve very large integers, because the chance of finding spurious
resonances is large. Therefore, we performed two searches, one limiting the integers to values
smaller than or equal to 5, and another one rising that limit to 10. Since the numbers of
the different kinds of resonant orbits are small, we bunched together the three different
models of each group and the results are presented in Table \ref{tab3} as percentages of the total
number of regular orbits. These results can be compared with those in the equivalent Tables from
\citet{AMNZ2007} and \citet{MNZ2009}. The first line gives the percentage of BBLs and the second
one the percentage of those that have no resonances in the search performed with integers up to 5,
i.e., those that can be regarded as boxes at that level of the search. The following lines give the
percentages of orbits which obey one single resonance and that have a percentage larger than 1 per
cent in, at least, one model; those with smaller percentages were bunched together in the line labelled
"Other". The line before the last one gives the percentages of orbits that obey two resonances and the last
line gives the percentages of those orbits for which no resonance was found in the search with integer
numbers up to 10. No resonant orbits with percentages larger than 1 per cent were found for integer numbers
larger than 5. Notice that the resonances found here are the same found by \citet{MNZ2009}, except for the
addition here of the (1,-2,1) resonance which barely made it to Table \ref{tab3} because its percentage is
1.07 in model E5.

As an example, Figure \ref{fig03} presents the x-y and x-z projections of orbit 1848 from
model E4a, which corresponds to the resonance (3,-2,0) type called fish.

\section{Conclusions}

An interesting result from Paper I is that the different models in each group gave essentially
the same results. The results of Table \ref{tab1} show the same for the percentages of the different
kinds of orbits: with very few exceptions (e.g., the percentages of OLATs in models E2b and E2c) the
differences among models of the same group fall within the $3 \sigma$ level. Due to
the small numbers involved, it is difficult to say if the same happens for the percentages of resonant
orbits, because in order to obtain the results of Table \ref{tab3} the three different models of
each group had to be bunched together. Nevertheless, at least the plots in Figure \ref{fig01} do
not show obvious discrepancies among the distributions of symbols that correspond to the different
models.

The analysis of the present results should be done bearing in mind that all these models are dominated
by chaotic orbits and that the regular orbits investigated here are just a minor component of the
orbital content: about 22 per cent of all the orbits in models E2 and E5 and only about 13 per cent of
the same in models E3 and E4. Thus, it is difficult to accept here the usual view that regular orbits
provide the framework for these models and chaotic orbits just help to fill in the gaps, actually it
seems to be the other way round.

The most abundant regular orbits turn out to be the SATs, in good agreement with the trend shown by a
comparison of the results of \citet{AMNZ2007} and of \citet{MNZ2009}: the fractions of SATs increase
several folds when going from non--cuspy to similar cuspy models. The percentages of SATS found here
are even larger than those of the cuspy models of \citet{MNZ2009}, probably because the present models
maintain the $\gamma \simeq 1$ slope down to their innermost regions (see Figure 3 of Paper I), while
the models of \citet{MNZ2009} show some tendency to flatten near the centre of the systems. 
The increase of the fractions of SATs as one goes from the E2 towards the E5 models is in agreement
with similar trends found by \citet{AMNZ2007} and \citet{MNZ2009}, respectivley for non--cuspy and
cuspy models.

The large fractions of OLATs in the E2 models is probably due to the fact that those models have
axial ratios $b/a \simeq c/a$ in their outer regions, as shown in Table \ref{tab2}.
For the other models, where the differences between the $b/a$ and $c/a$ ratios are larger, the fractions
of both ILATs and OLATs are very small, indeed. Similar trends were found by \citet{AMNZ2007} for
non--cuspy models and by \citet{MNZ2009} for cuspy ones.

The segregation of the orbits in energy bins shown in Table \ref{tab2} reveals some
interesting details, and one should bear in mind that orbital periods change enormously with energy, the
orbits in the lowest energy bin having periods of the order of several tenths of t.u. and those in the
highest of the order of several tens of t.u., i.e., a two orders of magnitude difference. The
concentration of the newly detected chaotic orbits in the highest energy bin may thus be in part
consequence of the use of a fixed time interval to compute the FT-LCNs in Paper I,
and of a fixed number of periods to compute the orbital frequencies in the present paper. Nevertheless
the fact that, in the sample of regular orbits from Paper I, those detected here as chaotic using a
longer integration interval also show preference for the higher energy bins, shows that the effect is
also in part real. In the case of sticky orbits, the effect is easy to understand because for the highest
energy bins a fixed integration interval would involve less "periods" of the chaotic orbit when it is
behaving more or less regularly and, thus, less chance to detect its truly chaotic nature. For the weakly 
chaotic orbits, instead, there is no obvious selection effect and they are probably more abundant at high
energies.

BBL orbits (recall that the bulk of them are not boxes, but boxlets) show a tendency to
occupy the higher energy bins as we go from the E2 to the E5 models. Even though they represented less
than one third of the regular orbits in all the models (and in some of them much less than that), they
are more than half of the regular orbits of the highest energy bin in models E4, and are almost as numerous
as the SATs in the same bin of models E3. SATs, in turn, are less numerous in the highest energy bin but,
except for the two mentioned cases, clearly dominate in all energy bins, as they do for all the energies
taken together. As could be expected, OLATs tend to concentrate in the higher energy bins and they manage
to outnumber the SATs in the highest energy bin of the E2 models. The opposite is true of the ILATs, but
the tendency is less pronounced, and they are even fairly abundant in the highest energy bin of the models
E2 where, as already indicated, those systems are close to being rotationally symmetric.

Most of the BBLs turn out to be resonant orbits. As could be expected from cuspy models, the fractions
of boxes are very small and clearly diminish even further when the search for resonances is extended to
larger integer numbers; it is worth recalling, however, that the bulk of the resonant orbits have
resonances with integers not larger than 5. The $(x,y)$ fishes, resonances (3,-2,0), are the most important
boxlets in models E3 and E4, in agreement with the results of \citet{MNZ2009}, but they are much less
abundant in models E2 and E5 and, in fact, neither boxes nor boxlets seem to be of much relevance to models
E2 and E5. Despite the differences between the fractions found here and those of \citet{MNZ2009}, it should 
be emphasized that the resonances whose fractions are larger than 1 per cent are essentially the same in
both investigations: only resonance (1,-2,1) was added here and it only exceeds 1 per cent in model E5.

\section{Acknowledgements}
We are very grateful to David Nesvorn\'y for the use of his code and to
R.E. Mart\'{\i}nez and H.R Viturro for their 
technical assistance. The suggestions of the referee, Luis
Aguilar, were very useful to improve the original manuscript and are gratefully 
acknowledged. This work was supported with grants
from the Consejo Nacional de Investigaciones Cient\'{\i}ficas y T\'ecnicas de la
Rep\'ublica Argentina, the Agencia Nacional de Promoci\'on Cient\'{\i}fica y
Tecnol\'ogica, The Universidad Nacional de La Plata and the Universidad Nacional de
Rosario.

\end{document}